\documentclass[%
 reprint,
superscriptaddress,
 amsmath,amssymb,
 aps,
]{revtex4-2}

\usepackage{graphicx}
\usepackage{dcolumn}
\usepackage{bm}
\usepackage{acronym}
\usepackage{amsmath}
\usepackage{amsfonts}
\usepackage{url}
\usepackage{amssymb}

\usepackage{pgfplots,filecontents}
\usepackage{hyperref}
\usepackage{cleveref}

\pgfplotsset{compat=newest}
\pgfplotsset{minor grid style = {dashed}}

\DeclareMathOperator*{\argmax}{arg\,max}

\begin{document}

\title{Machine-learning based noise characterization and correction on neutral atoms NISQ devices}

\author{Ettore Canonici}
\affiliation{Dept.\,of Physics and Astronomy \& European Laboratory for Non-Linear Spectroscopy (LENS), University of Florence, via G. Sansone 1, 50019 Sesto Fiorentino, Italy.}
\author{Stefano Martina}%
\affiliation{Dept.\,of Physics and Astronomy \& European Laboratory for Non-Linear Spectroscopy (LENS), University of Florence, via G. Sansone 1, 50019 Sesto Fiorentino, Italy.}
\author{Riccardo Mengoni}
\affiliation{CINECA, Via Magnanelli 6/3, 40033 Casalecchio di Reno, Bologna, Italy.}
\author{Daniele Ottaviani}
\affiliation{CINECA, Via Magnanelli 6/3, 40033 Casalecchio di Reno, Bologna, Italy.}
\author{Filippo Caruso}
\email{filippo.caruso@unifi.it}
\affiliation{Dept.\,of Physics and Astronomy \& European Laboratory for Non-Linear Spectroscopy (LENS), University of Florence, via G. Sansone 1, 50019 Sesto Fiorentino, Italy.}

\begin{abstract}
Neutral atoms devices represent a promising technology that uses optical tweezers to geometrically arrange atoms and modulated laser pulses to control the quantum states. A neutral atoms \acf{nisq} device is developed by Pasqal with rubidium atoms that will allow to work with up to 100 qubits. All \ac{nisq} devices are affected by noise that have an impact on the computations results. Therefore it is important to better understand and characterize the noise sources and possibly to correct them. Here, two approaches are proposed to characterize and correct noise parameters on neutral atoms \ac{nisq} devices. In particular the focus is on Pasqal devices and \ac{ml} techniques are adopted to pursue those objectives.
To characterize the noise parameters, several \ac{ml} models are trained, using as input only the measurements of the final quantum state of the atoms, to predict laser intensity fluctuation and waist, temperature and false positive and negative measurement rate. Moreover, an analysis is provided with the scaling on the number of atoms in the system and on the number of measurements used as input. Also, we compare on real data the values predicted with \ac{ml} with the a priori estimated parameters. Finally, a \acf{rl} framework is employed to design a pulse in order to correct the effect of the noise in the measurements. It is expected that the analysis performed in this work will be useful for a better understanding of the quantum dynamic in neutral atoms devices and for the widespread adoption of this class of \ac{nisq} devices.
\end{abstract}


\maketitle

\acrodef{nisq}[NISQ]{Noisy Intermediate Scale Quantum}
\acrodef{ml}[ML]{Machine Learning}
\acrodef{qml}[QML]{Quantum Machine Learning}
\acrodef{ai}[AI]{Artificial Intelligence}
\acrodef{qpu}[QPU]{Quantum Processing Unit}
\acrodef{svm}[SVM]{Support Vector Machine}
\acrodef{ann}[ANN]{Artificial Neural Network}
\acrodef{rnn}[RNN]{Recurrent Neural Network}
\acrodef{rl}[RL]{Reinforcement Learning}
\acrodef{mlp}[MLP]{Multi Layer Perceptron}
\acrodef{grf}[GRF]{Gibbs random field}
\acrodef{mae}[MAE]{Mean Absolute Error}
\acrodef{kl}[KL]{Kullback–Leibler}

\section{Introduction}
In the last few years we are witnessing a revolution in the field of quantum computing. The so called \acf{nisq} devices~\cite{preskill2018quantum} represent the state of the art in this field. The intermediate scale of such devices refers to the fact that at the best of our technologies, we are still capable of dealing with at most few hundreds of qubits. Several error correction codes have been developed to deal with such noise~\cite{cai2021bosonic,GoogleQuantumAI,lidar_brun_2013}, but they require the adoption of auxiliary qubits further decreasing the resources available for the computation.
\emph{Pasqal}~\cite{pasqal} has developed a \ac{nisq} device called \emph{Fresnel} based on a neutral atom quantum processor capable of using up to 100 qubits~\cite{Henriet2020quantumcomputing} and provides a \emph{Python} library called \emph{Pulser}~\cite{silverio2022pulser} that can be used to prepare a setting either to run it on the real machines or to simulate it on a built-in simulator. 

\acf{ml} is a field in the context of \ac{ai} that deals with the study and realization of models that learn to make predictions after being trained with data~\cite{shalev2014understanding,Bengio2016DeepLearning}. \acp{ann} are \ac{ml} methods organized in layers of artificial neurons that performs calculations with weighted summation of the inputs followed by non-linear activation functions. \ac{ml} methods has already developed in the context of quantum noise characterization~\cite{youssry2020characterization,martina2023machine,wise2021using,martina2022learning} and have already been adopted in the context of error estimation.
In~\cite{baireuther2018machine} the authors train a recurrent neural network to detect if certain errors happened in a quantum circuit and use the model to enhance a surface error correction code. Surface error correction codes allows an high error tolerance, however to be implemented they need an high number of physical qubits~\cite{fowler2012surface}. By contrast, in our proposed approach for noise mitigation, no additional qubits are needed for error detection. In fact, our purpose is to learn how to modify the pulses in such a way as to minimize the effect of noise without implementing error correction codes. Moreover, we estimate the noise in devices with the analog interface and not with the digital one. In fact, with neutral atoms devices it is possible to take advantage of analog and digital modes. With the former, laser pulses can be used to directly manipulate the Hamiltonian of the system: 
\begin{equation*}
    H = \frac{\hslash \Omega(t)}{2} \sum_{i} \sigma_{i}^{x} - \frac{\hslash \delta(t)}{2} \sum_{i} \sigma_{i}^{z} + \sum_{i<j} U_{ij}n_{i}n_{j}.
\end{equation*}
With the digital mode, on the other hand, it is possible to evolve the state of the system through quantum gates, thus creating quantum circuits.
 In~\cite{HarperQuantumNoiseEstimation} the authors consider the noise to have the form of a Pauli channel and make the assumption that the error rate is modeled with a \ac{grf}. Those assumptions allows the authors to effectively learn the parameters of the \ac{grf} to characterize the noise of a real IBM \ac{nisq} device. As discussed below, in our work we use a different noise formalization, in fact we resort on how the noise is implemented in the Pasqal simulator that we use to generate the data to train the deep learning model.

\ac{rl} is a \ac{ml} methodology that requires the presence of a simulator of an environment where an agent operates~\cite{sutton2018reinforcement}. The agent is usually implemented as a neural network that is trained to implement the policy that governs the actions of the agent. Initially, for each episode (the elementary phase of each \ac{rl} algorithm that is repeated over time and is constituted of a series of actions of the agent and reactions of the environment), the agent and the environment are initialized in some initial state. Then, the agent perceive some information about the environment and, based on that, the policy follows a probability distribution of the possible next actions that the agent can perform to change the state of the environment or the state of the agent within the environment. The episode continue with the choosing of the best action according to the policy and new steps until a predefined number of steps or some episode-ending condition. \ac{rl} have been already used in the context of state preparation and circuit optimization~\cite{RLNoiseMitigation,Goldschmidt2022modelpredictive,Porotti2022deepreinforcement}. In the context of noise correction, \ac{rl} have been adopted to correct the noise that degrades a state over time~\cite{sweke2020reinforcement} or to optimize existing quantum correction codes~\cite{colomer2020reinforcement}. In our work we instead focus on the task to correct the effects of the noise of a defined quantum dynamics without modifying the base pulse.
\begin{figure*}[!ht]
\begin{center}
\includegraphics[width=\textwidth]{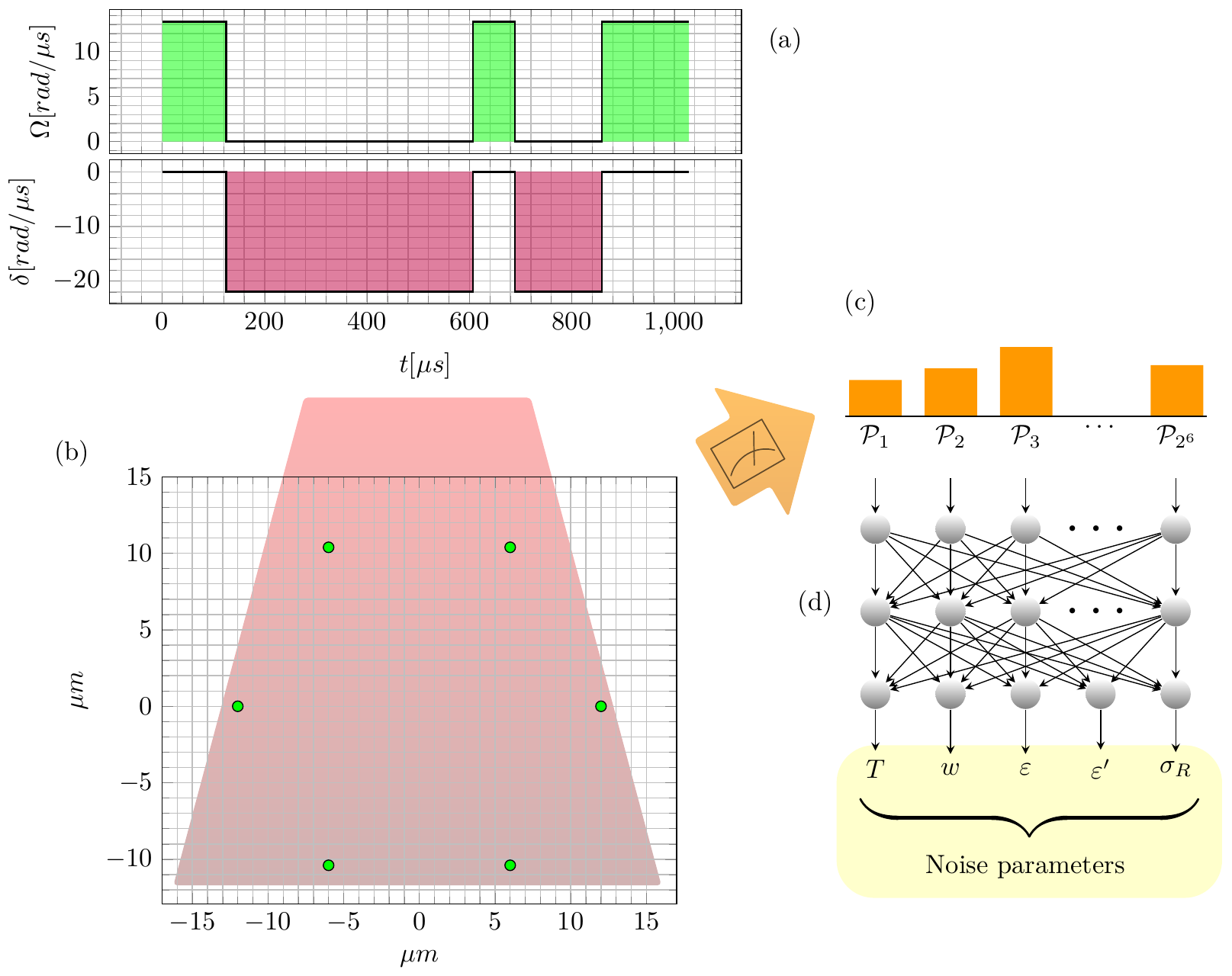}
\caption{Scheme of the noise estimation pipeline. A global pulse is defined by the shapes of Rabi frequency $\Omega$ and detuning $\delta$ (a). A register is prepared with the positions of a set of $n$ atoms (6 in the specific case) that are irradiated by the laser pulse (b). When the pulse ends, the excitation states of the atoms are measured and the process is repeated to gather statistics on the occupation probabilities $\bm{\mathcal{P}}=\mathcal{P}_1,\dots,\mathcal{P}_{2^n}$ (c). The probabilities are used as input to an \acf{ann} that predicts the noise parameters (d). The \ac{ann} is trained collecting a simulated dataset of probabilities labelled with the corresponding values of noise. The depicted setting is for the more general multiple parameters estimation. The difference for the single parameter estimation is that the neural network have only one output for $\sigma_R$ and the adopted pulses and atoms registers are different.}
\label{fig:intro}
\end{center}
\end{figure*}

\section{Noise benchmarking protocol}
A setting consists in the topological arrangement of atoms and the description of the laser pulses that interacts with them.
Then, the computation on \acp{qpu} is structured in cycles of three phases: (i) the preparation of the register, (ii) the quantum processing and (iii) the register readout. In particular, on neutral atoms devices, the preparation of the register is obtained using arrays of optical tweezers~\cite{schymik2020enhanced}. Initially the register is initialized with atoms in random positions and afterward the single atoms are moved in the desired positions. The quantum computation is performed analogically using laser pulses that interact with the register atoms and can excite them. The laser pulses are characterized by the values and shapes of the Rabi frequency $\Omega(t)$ and detuning $\delta(t)$. Finally, the register readout is performed by taking a fluorescence image to capture the energy levels of the atoms. In Pasqal \ac{nisq} devices, it is possible to prepare registers of maximum 100 atoms with a minimum distance of $4\mu m$ between them arranged in bidimensional structures in an area of maximum radius $50\mu m$.

\ac{nisq} devices, as the name suggests, are affected by several noise effects that limit their applicability and the operations that can be reliably executed on them. 
The devices
used in the realization of a quantum computer are not ideal, as they are affected by
noise: for example, lasers are not exactly monochromatic, and atoms are cooled by lasers to very low temperatures, but still non-zero. These imperfections have an impact by introducing errors during the preparation
of the system, its evolution over time, and the measurement. The effect is that the
measured probabilities of occupation are different from those we would have obtained
from an ideal environment. In general, there are different parameters that can be used to indicate different sources of noise in the device~\cite{de2018analysis}. In the present work we will focus on five parameters that are considered predominant for their effects: the laser intensity fluctuation $\sigma_R$ indicates the standard deviation of the fluctuation of the desired Rabi frequency of the laser pulse; the laser waist $w$ is the diameter of the gaussian laser beam; the false positive measurements $\varepsilon$ represents the probability of wrongly measure as excited an atom that was in the ground state; false negative measurements $\varepsilon'$ is the probability of measuring an excited atom in ground state.
\Cref{tab:errorsTable} shows those sources of noise and their estimated values provided informally by Pasqal. 
\begin{table}[ht]
\centering
\caption{Summary of the main noise parameters with their respective values. We considered the parameters that are expected to have a predominant effect.}
\begin{tabular}{c|c|c}
\multicolumn{1}{c}{\textbf{Description}} & \multicolumn{1}{c}{\textbf{Parameter}} & \multicolumn{1}{c}{\textbf{Value}}\\
\hline
Laser Intensity fluctuation & $\sigma_R$ & $3\%$ \\
Laser waist & $w$ & $68 \mu m$ \\
Temperature & $T$ & $30\mu K$ \\
False positive measurement & $\varepsilon$ & $3\%$ \\
False negative measurement & $\varepsilon '$ & $8\%$ \\
\hline
\end{tabular}
\label{tab:errorsTable}
\end{table}

The objective of our work is the implementation of \ac{ml} models to: 
(i) provide a quantitative estimate of the noise; (ii) mitigate the effects of the noise. 
We decided to formulate a supervised regression task to quantitatively estimate the noise~\cite{HarperQuantumNoiseEstimation} and to use a \acf{rl} framework~\cite{sutton2018reinforcement} to mitigate the noise effect. Regarding the noise characterization, our aim is to show that it is possible to estimate the noise parameters in the form of mean values and error intervals.
As depicted in \cref{fig:intro}, the workflow begins with the simulation of various executions, with different noise parameters, of a quantum dynamic where a global pulse irradiates all the $n$ atoms of a register. Afterward, the atoms occupation probabilities, that we call $\bm{\mathcal{P}} = \mathcal{P}_1,\dots,\mathcal{P}_{2^n}$, are collected and used to train \ac{ann} models to predict the noise parameters that were used to perturb the dynamics: \emph{temperature}, laser \emph{waist}, false positive measurement rate \emph{$\varepsilon$}, false negative measurement rate \emph{$\varepsilon'$} and intensity fluctuation \emph{$\sigma_R$}.
At the end, the trained models are used on prediction with the real data, obtaining an estimation of the noise parameters. For the the simulations used in the generation of the data and for the training of the models, we use our servers with Nvidia TITAN RTX and GeForce RTX 3090 GPUs. Moreover we could also make use of the CINECA Marconi100 supercomputer.

The rest of the paper is structured as in the following. First, in \cref{subsec:singleParameterCharacterization} we consider the simpler problem of characterizing only a single noise parameter, then in \cref{subsec:multipleParametersCharacterization} we show the results of the characterization of all the aforementioned parameters. In \cref{sec:errorCorrection} we illustrate the \ac{rl} error correction protocol that we adopt.

\section{Noise characterization}
\subsection{Single parameter scenario}\label{subsec:singleParameterCharacterization}
In this section we consider the estimation of a single noise parameter. After preliminary analysis, we decided to focus on the noise effects that comes from the laser intensity fluctuations $\sigma_R$.

Before describing the used methods, let us introduce the notation. We will denote by $s_i$ the system composed of $i$ qubits. Globally, we consider systems with a number of qubits from 2 to 5 and in the case of 4-qubit systems we denote 6 different topologies with an extra alphanumeric index from $a$ to $f$. Specifically, $s_{4a},s_{4b},\dots,s_{4f}$.
Globally, we collected the measurements of nine different runs on the real Pasqal \ac{nisq} devices (6 different topologies with 4 atoms and single topologies with 2,3 and 5 atoms) characterized by a pulse with constant Rabi frequency $2\,\pi\,rad/\mu s$ of duration $660\,ns$ and null detuning but with different number and positions of the atoms. 

In order to train the \ac{ml} models to predict the values of $\sigma_{R}$, we simulate the data for computation on the nine registers with different amount of simulated noise effects.
In detail, we preliminarily generate a sequence of $10\,000$ $\sigma_{R}$ values extracted from a uniform distribution $\mathcal{U}(0,0.15)$. These values are used to add noise in an equal number of simulations, whose results are occupation probability vectors. Therefore, in the end, $10\,000$ samples are obtained.
This procedure is repeated for each of the 9 quantum systems we mentioned above. The occupation probabilities associated with the corresponding values of $\sigma_R$ for the 9 systems are used to evaluate two different scalings: (i) in the quantum register size comparing increasingly larger systems of 2,3,4 and 5 qubits and (ii) in the number of measurements of multiple systems with 4 qubits where the occupation probabilities of all the systems simulated with the same values of $\sigma_R$ contributes to gather information on the noise effects during the training of the \ac{ml} models. In detail, we decided to use as input to the \ac{ml} models the concatenation of the probabilities of the systems and, for two systems $s_A$ and $s_B$, we indicate the latter with the notation $s_A\oplus s_B=\mathcal{P}_{1,A},\dots,\mathcal{P}_{2^n,A},\allowbreak \mathcal{P}_{1, B},\dots,\mathcal{P}_{2^n, B}$.
In both scaling, the procedure is always the same: 20 models are trained on each dataset through a 20-fold cross validation. 
From the 20 predicted parameter values, the average value and the standard deviation can be obtained to include the variability of the models' predictions.
Both analyses are performed with linear regression as baseline model and with \ac{ann}s.
Regarding \ac{ann}s, they are trained for 150 epochs with the Adam optimizer and with hyperparameter optimization. For a more in-depth discussion of the technical details related to model design and hyperparameters optimization reefer to \cref{sec:singleParameterCharacterizationAPPENDIX}.

\begin{figure*}[!ht]
\centering
\includegraphics[width=\textwidth]{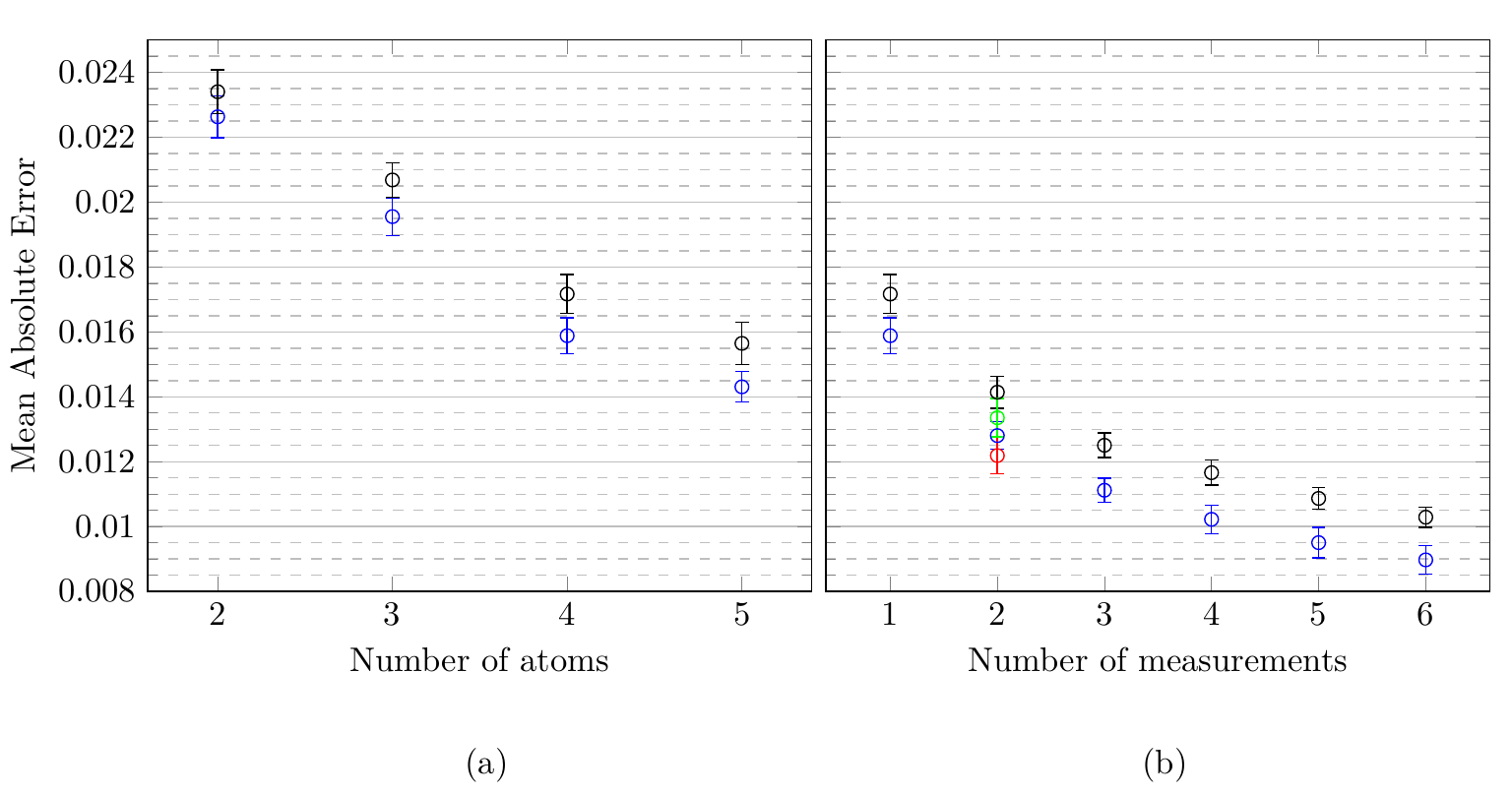}
\caption{Scaling of single measurement for systems with an increasing number of atoms (a) and scaling in the number of measurements for systems with four atoms (b). We report the average absolute errors and standard deviations for 20 linear regression (in black and green) and 20 \ac{ann} (in blue and red) models in the predictions of $\sigma_R$ on the synthetic validation set. The models in (a) uses as input the measurements of $s_2$, $s_3$, $s_{4a}$ and $s_5$. 
The models in (b) uses as input one or more concatenated measurements of runs of the settings with four atoms (the fourth pair of points in (a) is equal to the first pair in (b)). Indicating with $\cdot\oplus\cdot$ the concatenation of the measurements of the settings, we report in (b) in black and blue ${s_{4a},}\, {s_{4a}\oplus s_{4c},}\, {s_{4a}\oplus s_{4c}\oplus s_{4d},}\, {s_{4a}\oplus s_{4c}\oplus s_{4d}\oplus s_{4e},}\, {s_{4a}\oplus s_{4c}\oplus s_{4d}\oplus s_{4e}\oplus s_{4f}}$ and in green and red ${s_{4a}\oplus s_{4b}}$.}
\label{fig:scaling}
\end{figure*}

\begin{figure*}[!ht]
\centering
\includegraphics[width=\textwidth]{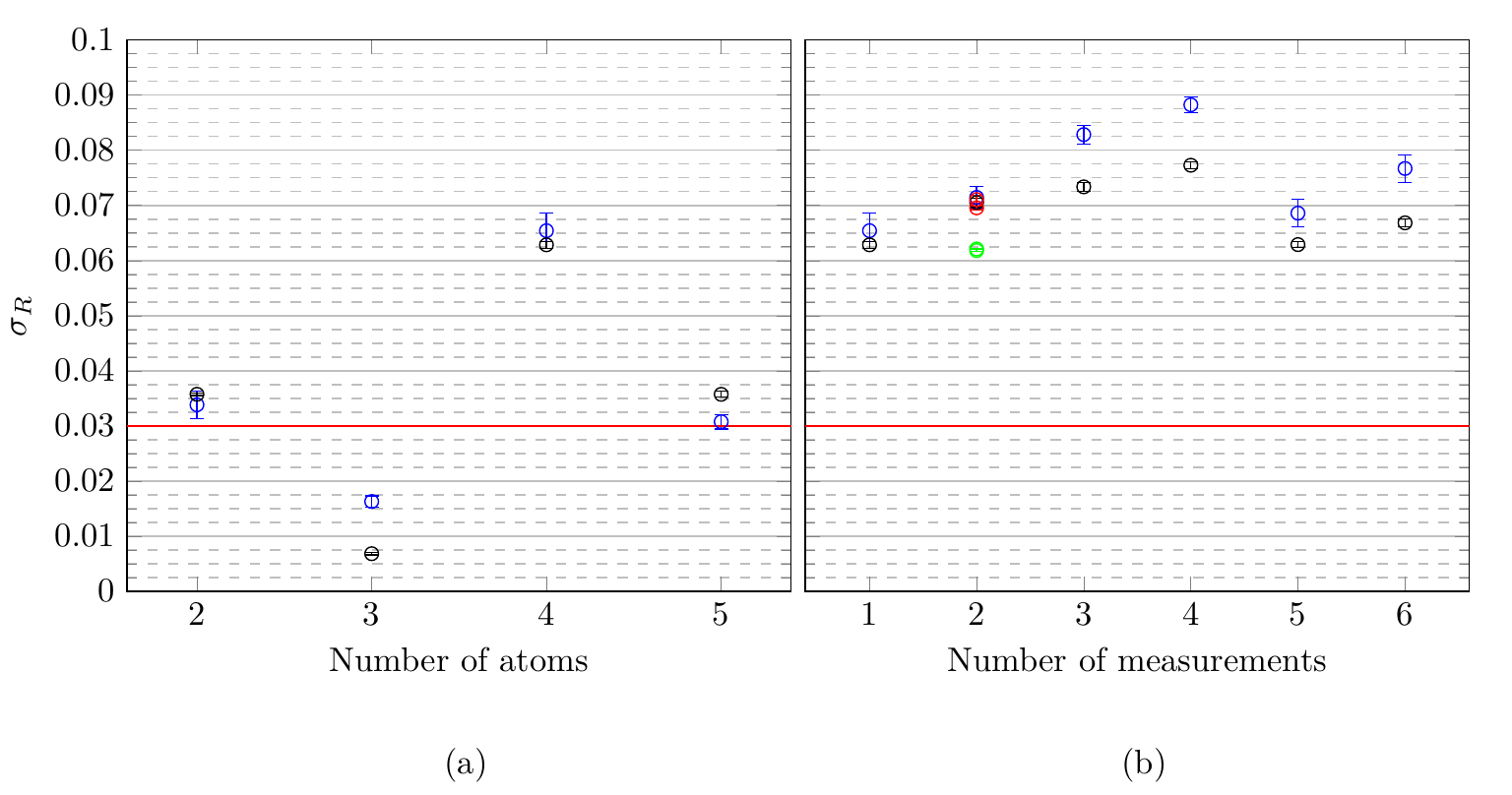}
\caption{Predictions on real data of the value of $\sigma_R$ for the models trained for the scaling in the number of atoms (a) and in the number of measurements (b) reported in \cref{fig:scaling}. We report the average values and standard deviations for the 20 linear regression (in black and green) and the 20 \ac{ann} (in blue and red) models in the predictions of $\sigma_R$ using a set of real measurements of the settings described in \cref{tab:registers} run on the Pasqal \ac{nisq} devices. The models in (a) uses as input the measurements of $s_2$, $s_3$, $s_{4a}$ and $s_5$. 
The models in (b) uses as input one or more concatenated measurements of runs of the settings with four atoms (the fourth pair of points in (a) is equal to the first pair in (b)). 
We report in (b) in black and blue the incremental concatenation of $s_{4a}$, $s_{4c}$, $s_{4d}$, $s_{4e}$ and $s_{4f}$. In green and red we report the concatenation of $s_{4a}$ and $s_{4b}$. The order of the real measurements for the latter concatenation is irrelevant, thus we report two green and two red points (almost overlapping and not clearly discernible) to consider the two possible concatenations. The horizontal red line indicates the value of $3\%$ for $\sigma_R$ estimated by Pasqal.}
\label{fig:predictions}
\end{figure*}

In the following, the \ac{ml} models are trained and validated on the simulated data, and subsequently they are also tested on the real measurements. Using the simulated validation data, it is possible to monitor how the model is capable of generalization to unseen measurements. In this regard, we report in \cref{fig:scaling} (the scaling (i) in \cref{fig:scaling}(a) and (ii) in \cref{fig:scaling}(b)) the \ac{mae}, averaged for all the samples of the validation set, between the predicted values of $\sigma_R$ and the ground truth that we recall is the value of $\sigma_R$ used to perform the simulation.
Again, having 20 estimates (one for each model), we calculate mean value and standard deviation of the \ac{mae} to provide more robust results with associated uncertainty.
Regarding the estimation on the real data, we show in \cref{fig:predictions} (\cref{fig:predictions}(a) for the scaling (i) and \cref{fig:predictions}(b) for the scaling (ii)) the mean values and standard deviations along the 20 models of the predicted values of $\sigma_R$.
In both \cref{fig:scaling} and \cref{fig:predictions}, the result of the training of linear regression models are depicted in black and the results of \ac{ann} in blue. 
Additionally, in \cref{fig:scaling}(b) and \cref{fig:predictions}(b) we highlight in green for the linear regression and in red for the \ac{ann} a specific case: the concatenation of the measurements on two peculiar settings with four atoms, $s_{4a}$ and $s_{4b}$, that have not only the same amount of atoms but also exactly the same topology. Therefore, the latter can be seen as a special case of the scaling (ii) where multiple measures of the same system are performed. Moreover, for the real measurements we consider both orderings $s_{4a}\oplus s_{4b}$ and $s_{4b}\oplus s_{4a}$ whose prediction results are reported with two couples of green and red points in \cref{fig:predictions}(b) (not clearly visibles in the plot because they are almost overlapping).

As expected, the prediction error is decreasing with the number of atoms in the system because we get more information on the dynamic and thus on the noise influencing it. In \cref{fig:scaling} we can also observe that \ac{ann} are in general more powerful respect to linear regression models (at the cost of more resource-intensive computations). In fact, the errors for the \ac{ann} models are always lower respect to the errors of linear regression models and the difference is more pronounced increasing the number of atoms and measurements. This can be explained with a better capacity of \acp{ann} to model complex dynamics.

Overall, comparing \cref{fig:scaling}(a) with 5 atoms and \cref{fig:scaling}(b) with number of measurements equal to 2, seems to be more convenient to consider more measurements respect to increase the number of atoms of the setting. Also, comparing the green and red points with the black and blue ones for the same number of measurements in \cref{fig:scaling}(b), can observe that can be slightly better to consider multiple measurements of the same setting with the same topology respect to collect measurements of a different setting with the same number of atoms.

We observe in \cref{fig:predictions} that the values of $\sigma_R$ predicted for the measurements of settings with $2$ and $5$ atoms are close to the estimated value of $3\%$, however the prediction for the setting of $3$ atoms is lower and the predictions for all the settings with $4$ atoms, and concatenation of them, are around $7\%$. An explanation for this mismatch can be that
the real data used for the experiments was collected when the device was still under development. 
Moreover the predictions consider only $\sigma_R$ as a variable source for the noise, thus variations of the other noise parameters in the real machine influence the predictions of $\sigma_R$. Nevertheless, it is remarkable that the trained models have low standard deviations for the predictions that, even if this does not exclude an high bias error, still suggest a low variance error for the models. We can also observe that the order of the measurement for the settings $s_{4a}$ and $s_{4b}$ do not influence the predicted values -- in fact the two green circles and the two red circles in \cref{fig:predictions} are almost overlapping.

To summarize, noise estimation based on supervised learning is possible. The protocol we presented seems to suggest merging data from multiple similar registers instead of larger registers directly. This may be useful because of the difficulty in simulating larger systems. In addition, the estimates obtained are derived by averaging estimates from 20 models. Moreover, the associated standard deviation is small relative to the predicted value, so all 20 models converge to very similar values.
Finally, we repeat that having neglected several noise sources, the parameter values found could be effective values.

\subsection{Multiple parameters characterization}\label{subsec:multipleParametersCharacterization}
In this section we train a deep learning model in a multioutput regression setting to estimate the values of all the noise parameters in \cref{tab:errorsTable}.
We simulated a dataset of $54\,000$ labelled samples for the 6-qubit system whose topology can be observed in \cref{fig:intro}(b).
The used pulse sequence that defines the dynamics is shown in \cref{fig:intro}(a).
Analogously to the scaling experiments in the previous section, the measurement for each simulation is obtained sampling 500 runs. The values used in the simulations for each parameter are: 
$\sigma_R = \mathcal{U}(0, 0.15)$, $w (\mu m) = \mathcal{U}(0, 200)$, $T (\mu K) = \mathcal{U}(0, 100)$, $\varepsilon = \mathcal{U}(0, 0.15)$ and $\varepsilon' = \mathcal{U}(0, 0.15)$.

After finding the best set of hyper-parameters, 20 models are trained using the cross validation procedure to exploit the entire dataset and to obtain the standard deviations of the predictions.
Each one of the 20 models is trained with early stopping for a maximum of 150 epochs.
Further technical details related to \acp{ann} design and hyperparameter optimization can be found in \cref{sec:multipleParametersCharacterizationAPPENDIX}.
\begin{table}[ht]
    \centering
    \caption{Predicted values on real data expressed as average and standard deviation of 20 models trained on cross validation. The last column report for practicity the same estimated values of \cref{tab:errorsTable}.}
    \label{tab:results}
    \begin{tabular}{r|rcl|c}
         \multicolumn{1}{c}{\textbf{Parameter}} & \multicolumn{3}{c}{\textbf{Predicted value}} & \multicolumn{1}{c}{\textbf{Estimated value}}  \\
         \hline
         $\sigma_R$ & $0.079$ & $\pm$ & $0.005$ & $0.03$\\
         $w$ & $122\mu m$ & $\pm$ & $6$ & $68\mu m$\\
         $T$ & $56\mu K$ & $\pm$ & $4$ & $30\mu K$\\
         $\varepsilon$ & $0.082$ & $\pm$ & $0.010$ & $0.03$ \\
         $\varepsilon'$ & $0.078$ & $\pm$ & $0.005$ & $0.08$ \\
         \hline
    \end{tabular}
\end{table}

In \cref{tab:results} we show the resulting estimation of the main noise factors. Each reported value is the average of the 20 models trained on different splits with the corresponding standard deviation.
We observe that the predicted values do not match those estimated by Pasqal, although all 20 models always converge to very similar values of the predictions. In this regard, the same considerations expressed at the end of \cref{subsec:singleParameterCharacterization} are also valid for multi-parameter estimation: ie, that the parameter predictions obtained could be effective values that incorporate other neglected effects (noise sources, influence of other neighboring atoms, etc.).
Another possible factor could be that the measurements came from a prototype NISQ, just as in the case of those used in \cref{subsec:singleParameterCharacterization}. Therefore, we can expect more agreement in the future as a result of technical improvements.
Moreover, it is worth noting that, even if for the experiments in this section the setting and the pulse are different to the ones used in \cref{subsec:singleParameterCharacterization}, the predicted value for $\sigma_R$ is comparable to the ones obtained for the estimation of the same parameter in the settings with four atoms previously illustrated. 

\section{Error correction}\label{sec:errorCorrection}
Many techniques have been developed in the theory of classical error-correcting codes~\cite{hamming1950error,peterson1972error}. The key idea on which they are based is mainly redundancy. Nonetheless, the addition of redundancy is not immediate in \ac{nisq} devices because of the no cloning theorem~\cite{wootters1982single}. However, some sort of redundancy can be achieved in quantum devices by expanding the system to more qubits~\cite{shor1995scheme}. In fact, all the most used quantum error correction techniques require the use of more qubits than the ones strictly necessary for the computation~\cite{roffe2019quantum} but it is not feasible with NISQ devices. Therefore,
we propose to verify that it is possible to mitigate the effects of quantum noise without extra qubits through the use of \ac{rl} techniques.
\ac{rl} is a \ac{ml} area where an agent learns which actions to perform in order to maximize a reward~\cite{sutton2018reinforcement}. Schematically, we can say that this is a closed-loop problem because the actions of the learning system influence subsequent inputs. In addition, the learner does not know a priori which action to perform and has to find out for himself through trials and errors which actions lead to larger rewards. Actions can influence not only the immediate reward but also future rewards. 
\ac{rl}, unlike Supervised Learning, does not require labelled input-output pairs, but focuses on finding a balance between exploration of the actions space in an environment and exploitation of the acquired knowledge. The agent must exploit what it already knows in order to obtain reward, but it must also explore in order to make better action selections in the future. The trade-off is that neither exploration nor exploitation can be exclusively pursued without failing in the task. The agent must try a variety of actions and progressively favour those that seem to be the best.
Any problem of learning goal-oriented behaviour can be reduced to three signals that are exchanged between an agent and its environment: a signal to represent the choices made by the agent (the actions), a signal to represent the basis on which the choices are made (the states) and a signal to define the agent's goal (the rewards). In detail, for each action of the agent at time $t$, its effects on the environment are quantified by a reward $r_t$. Then the objective of the training is to maximize the discounted cumulative reward $R_{t_0}=\sum_{t=t_0}^\infty \gamma^{t-t_0}r_t$, where the discount $\gamma\in(0,1)$ is an hyperparameter that controls the importance of rewards far in the future respect to the ones immediately after $t_0$. This objective is implemented with the idea that if we would have a function $Q^*:State\times Action\rightarrow \mathbb{R}$ that given a state and an action performed over that state, returns the cumulative discounted reward, then the policy can be implemented with $\pi^*(s)=\argmax_a Q^*(s,a)$. In general, $Q^*$ is unknown and is approximated by a neural network. For a defined policy $\pi$, the $Q$ function obeys the Bellman equation $Q^\pi (s,a)=r+\gamma Q^\pi(s',\pi(s'))$ where $r$ and $s'$ are respectively the reward and the next state obtained after the action $a$ on the state $s$. The neural network that defines $Q$, and then the agent, is trained minimizing over a batch of transitions the Huber loss $\mathcal{L}(\delta)$ of the temporal difference error $\delta=Q(s,a)-(r+\gamma\max_a Q(s',a))$.

\begin{figure*}[!ht]
      \centering
      \includegraphics[width=0.9\textwidth]{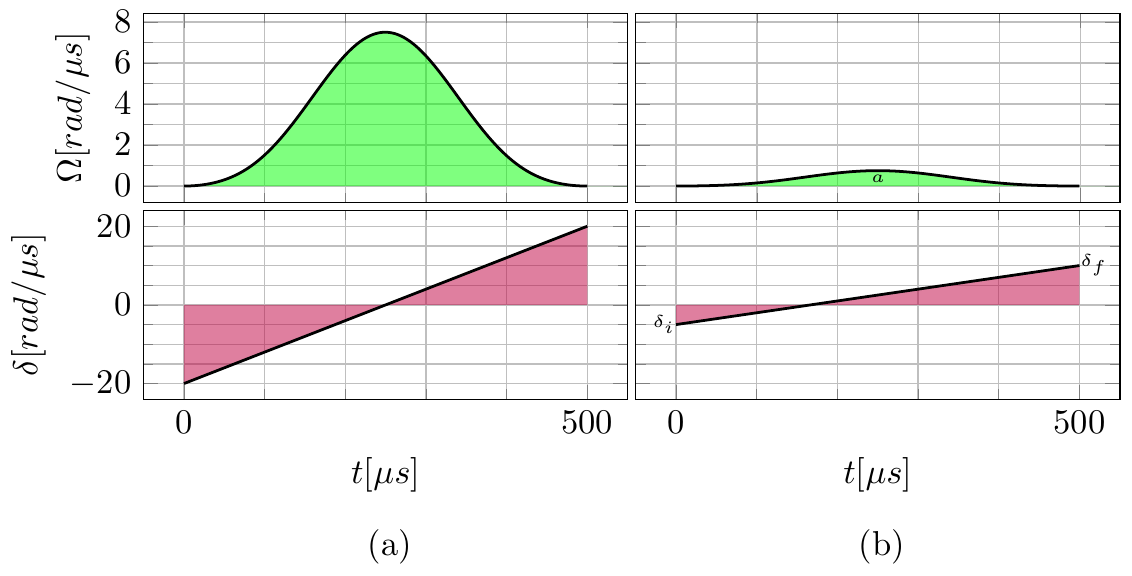}
      \caption{Standard pulse $P$ (a) to be corrected with a correction pulse $P'$ (b) to be added after $P$ to counteract the effects of the noise. The Rabi frequency $\Omega$ is depicted in green and the detuning $\delta$ in purple. 
      $P$ is a pulse of duration $T=500 ns$, Gaussian Rabi profile with area equal to $\pi/2$ and detuning in the form of a ramp from $\delta_0 = -20$ rad/$\mu s$ and $\delta_T = 20$ rad/$\mu s$.  $P'$ is a pulse with the same duration and characteristics of $P$ but with variable Rabi area $a$, initial detuning $\delta_i$ and final detuning $\delta_f$.}
      \label{fig:pulseToCorrect}
\end{figure*}
We choose to correct the standard impulse $P$ depicted in \cref{fig:pulseToCorrect}(a) applied to a single qubit. $P$ has a Gaussian profile in the Rabi frequency $\Omega$ of duration $T=500$ $ns$ and area $\pi/2$ and a ramp profile in detuning $\delta$ of duration $T=500$ $ns$ with $\delta_0 = -20$ rad/$\mu s$ and $\delta_T = 20$ rad/$\mu s$. The choosen approach to correct the noise is to apply the correction pulse  \cref{fig:pulseToCorrect}(b) to be placed after the pulse to be corrected and having the same characteristics and length of $T=500$ $ns$. In detail, we choose a Gaussian profile in the Rabi frequency with variable area $a$ and a ramp profile in detuning $\delta$ with variable initial $\delta_i$ and final $\delta_f$.
In such a way, the final atoms occupation probabilities with the application of the corrected pulse
$\bm{\mathcal{P}}_{P+P'}^{noisy}$ and after the ideal pulse $\bm{\mathcal{P}}_P^{ideal}$ are \emph{closer} than $\bm{\mathcal{P}}_P^{noisy}$ and $\bm{\mathcal{P}}_P^{ideal}$.
By the notation $\bm{\mathcal{P}}_{j}^{i}$ we denote the measurement $\bm{\mathcal{P}}$ obtained after running a simulation with the pulse $j$ with or without noise (respectively, $i=noisy$ or $i=ideal$).
 The training allows to find the three optimal parameters $a$, $\delta_i$ and $\delta_f$
 for the correction impulse $P'$. 

\begin{figure*}[!ht]
\centering
\includegraphics[width=0.8\textwidth]{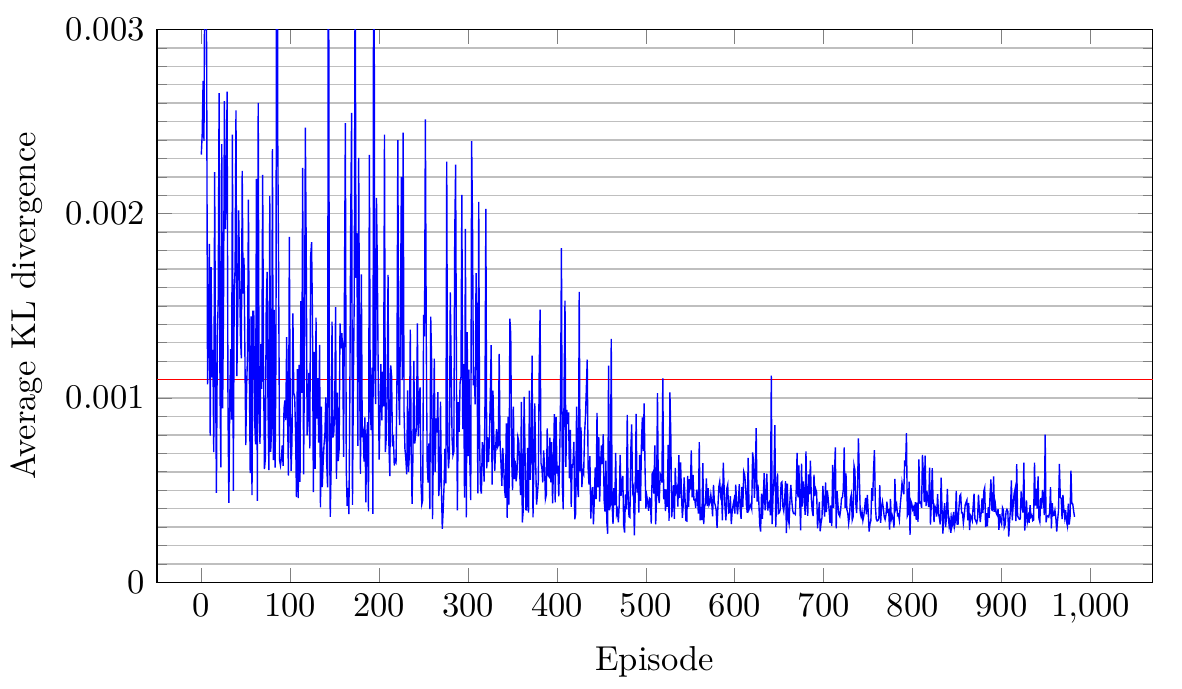}
\caption{Evolution of the \ac{kl} divergence between the corrected noisy simulation and the ideal one averaged for each episode. The red line is the reference value of $0.0011$ for the \ac{kl} divergence between the uncorrected noisy simulation and the ideal one averaged over $100$ simulations.}
\label{fig:KLdivReward}
\end{figure*}

In our \ac{rl} framework, the state is represented by the occupation probabilities that are estimated from 
the average of $10$ independent noisy simulations whose probabilities are extracted from the amplitudes of 25 quantum states uniformly sampled along the simulated dynamic. At the beginning of each episode we choose $a=\pi/20$ and $\delta_i=\delta_f=0$ and they can have values in the ranges $a\in[0,\pi/2]$ and $\delta_i,\delta_f\in[-20,20]$.
The agent, implemented with an \ac{ann} that have an input layer of 50 units (2 basis for each one of the 25 intermediate states), two ReLU hidden layer of 128 neurons and an output layer of 6 neurons, selects one among four possible actions: $a^t = a^{t-1} + \Delta a$, $a^t = a^{t-1} - \Delta a$, $\delta_i^t = \delta_i^{t-1} + \Delta\delta_i$, $\delta_i^t = \delta_i^{t-1} - \Delta\delta_i$, $\delta_f^t = \delta_f^{t-1} + \Delta\delta_f$, $\delta_f^t = \delta_f^{t-1} - \Delta\delta_f$. We choose fixed values for $\Delta a = \pi/200$  and  $\Delta\delta_i=\Delta\delta_f = 0.2$.
Each episode is constituted of a series of steps at increasing values of $t$. For each step, the chosen action is applied, a correction impulse $P_t'$ characterized by $a^t$, $\delta_0^t$ and $\delta_f^t$ is generated and used in a new simulation obtaining a new probability vector $\bm{\mathcal{P}}_{P+P_t'}^{noisy}$ for the final quantum state of the corrected noisy simulation and the reward $r(t)$ before proceeding with the next step. The episode ends when the action causes $a$, $\delta_0$ or $\delta_f$ to go out of boundaries or after 100 steps. The reward is defined as: 
\begin{equation}
    r(t) = \begin{cases} 
      1 & \text{if }\quad \left|\bm{\mathcal{P}}_{P+P_t'}^{noisy} - \bm{\mathcal{P}}_{P}^{ideal}\right|_1 \ < \ \left|\bm{\mathcal{P}}_{P+P_{t-1}'}^{noisy} - \bm{\mathcal{P}}_{P}^{ideal}\right|_1, \\
      0 &  \text{otherwise},
   \end{cases}
\end{equation}
where $|\cdot|_1$ is the $\ell_1$ norm. Specifically, the reward is $1$ if the last action at step $t$ makes the corrected noisy simulation closer to the ideal one respect to the previous step $t-1$ and $0$ otherwise. During the training we monitor
the \ac{kl} divergence between $\bm{\mathcal{P}}_{P+P_t'}^{noisy}$ and $\bm{\mathcal{P}}_P^{ideal}$:
\begin{equation}
    D_{KL}(\bm{\mathcal{P}}_{P+P_t'}^{noisy}, \bm{\mathcal{P}}_P^{ideal}) = \sum_{i=1}^2 \left(\bm{\mathcal{P}}_{P+P_t'}^{noisy}\right)_i\log\left(\frac{\left(\bm{\mathcal{P}}_{P+P_t'}^{noisy}\right)_i}{\left(\bm{\mathcal{P}}_P^{ideal}\right)_i}\right),
\end{equation}
averaged for all the steps $t$ within each episode. The evolution of the averaged \ac{kl} divergence for the $1\,000$ training episodes is reported in \cref{fig:KLdivReward} where we can observe that it effectively decreases below the reference value of $D_{KL}(\bm{\mathcal{P}}_P^{noisy}, \bm{\mathcal{P}}_P^{ideal})=0.0011$ reported with the red line and calculated with the average for $100$ noisy simulations without the correction pulse.

\section{Conclusions and outlooks}\label{sec:summary}
We presented two applications of \ac{ml} to the context of quantum noise characterization and correction. To characterize the noise we collected a dataset of multiple simulated noisy measurement of different settings in Pasqal quantum machines to train \ac{ml} models and we test them on real data. For the noise correction we trained a \ac{rl} model to find a correction pulse to counteract the effects of the noise affecting a simulated test setting. Regarding the noise characterization, we compared \ac{ann} with linear regression models in predicting the value of the laser intensity fluctuation $\sigma_R$, scaling the number of qubit in the register and the number of measurements of the system. We found that \ac{ann} perform better than linear regression and that the model accuracies increases both with the number of qubits and with the number of measurements. Moreover, we have insights that in order to better characterize the noise parameters it is more effective to increase the number of measurements respect to the number of qubits. When we tried to predict the noise parameters on real \ac{nisq} devices we found that, for every set of measurement, 40 different models (\ac{ann} and linear regression trained independently in a 20 fold cross validation setting) agree on the predictions and therefore the variance error is low. Finally, we trained 20 \ac{ann} models in a multiregression setting to predict five different noise parameter values and also in this case the models agree between them when tested on real data.
Regarding the noise correction, the proposed approach successfully learns to correct a simulated noisy pulse and to make the measured probabilities closer to the ideal ones.

We believe that the results presented in this work can be used to better quantify the effects of the noise affecting the Pasqal, and in general neutral atoms, \ac{nisq} devices and to counteract those effects. The presented tecniques are dependent on the atoms topology and the pulse shape. 
Thus, the \ac{ml} models can be trained to characterize and correct the noise of single quantum gates that compose more complex Hamiltonians. 

The accuracy of the predicted noise parameters depends on the accuracy of the simulation and in particular on the accuracy of the simulator noise model. 

In previous works~\cite{martina2023machine,martina2023deep,martina2022learning,fanchini2021estimating} and in preliminary experiments using Pasqal simulator, there is an evidence of the improvement of the noise characterization when more temporal statistics are collected. We adopted this strategy in this paper for the noise correction, where the occupation probabilities are obtained from the amplitudes of the intermediate quantum states sampled at regular steps within the simulated dynamic. However, in real \ac{nisq} devices, intermediate measurements of the dynamic are less straightforward because of the impossibility of observing a system without changing it. We can obtain the same effect independently measuring incremental subdynamics from $t=0$ to subsequent time steps of the full dynamic.
To implement this approach on Pasqal machines, we can design a full pulse that is subsequently split in sub-pulses at times $[t_0,t_1],[t_1,t_2],\dots,[t_{n-1},t_n]$. The measurements at time $t_k$ for $k=1,\dots,n$ can be obtained initialising the register always to the same initial setting and performing the computation considering the effects of all the sub-pulses spanning the times $[t_0,t_k]$ from the first to the one before $t_k$. The \ac{ml} models can then process all the measurements obtained at times $t_1,\dots,t_k$ and in that way we expect to obtain better results for the characterization of the noise. Moreover, we can also use \ac{ann} more suitable for data organized in temporal sequences, i.e. \ac{rnn}.

Finally, in the context of \ac{qml}~\cite{dunjko2016quantum,aimeur2006machine} our work is framed as a classical \ac{ml} approach to process quantum data. Future research lines may include the design of \ac{qml} models for the noise characterization and correction implemented directly within the quantum dynamic of neutral atoms devices or of other \ac{nisq} devices. For instance pattern matching \ac{qml} techniques~\cite{das2023quantum} can be adapted for the identification of noise patterns~\cite{martina2022learning} characteristics to the neutral atoms dynamics.

\medskip
\textbf{Acknowledgements} \par 
This work was financially supported by the European Union’s Horizon 2020 research and innovation programme under FET-OPEN GA n.\,828946--PATHOS. We acknowledge the CINECA award under the ISCRA initiative, for the availability of high performance computing resources, as \emph{Marconi100} supercomputer, and their support. S.M. acknowledges financial support from PNRR MUR project PE0000023-NQSTI. Finally, we are also thankful to Pasqal for the provided data that we have used to test our protocol.

\medskip

%
\bibliography{mainArxiv}








\section{Methods}
\subsection{Single parameter characterization}\label{sec:singleParameterCharacterizationAPPENDIX}
In this section we describe with more details the topologies of the analyzed quantum systems and the \ac{ann} models used in \cref{subsec:singleParameterCharacterization}.
\begin{table}[ht]
\centering
\caption{
Quantum systems used for single parameter estimation $\sigma_R$. By the notation $s_i$ we denote the system formed by $i$ atoms. In the case of 4 atoms, having used 6 different systems for the spatial arrangement of atoms, we use an additional subscript $s_{4j}$, with $j=\{a,b,c,d,e,f\}$.
Also, when a quantum register of a system $s_k$ is entirely contained in the quantum register of a larger system $s_k'$, with $k' > k$, we use the notation $s_k \subset s_{k'}$.}
\label{tab:registers}
\begin{tabular}{c|ccccccccc|}
     \textbf{Name} & $s_2$ & $s_3$ & $s_{4a}$ & $s_{5}$ & $s_{4b}$ & $s_{4c}$ & $s_{4d}$ & $s_{4e}$ & $s_{4f}$\\
     \cline{2-10}
     \textbf{Atoms} & $2$ & $3$ & $4$ & $5$ & $4$ & $4$ & $4$ & $4$ & $4$\\
     \cline{2-10}
     \textbf{Properties} & $\subset s_{4a}$ & $\subset s_5$ & $\subset s_5$ & & $= s_{4a}$ & & & & \\
\end{tabular}
\end{table}
The registers summarized in \cref{tab:registers} have an incremental number of atoms from 2 to 5 and some of them are chosen in a way such that the positions of the atoms of every register are included in the subsequent ones as far as possible.

To be precise, with the notation $s_k \subset s_{k'}$ and $k<k'$, we indicate that the quantum register of $s_{k'}$ is the same as that of $s_k$ with the addition of an atom and that therefore the coordinates of the atoms in common are the same.
In detail, the setting with five atoms ($s_5$) have atoms in the same positions of the ones of the settings of dimensionality four ($s_{4a}$) and three ($s_3$) plus extra atoms in other positions. Moreover, $s_{4a}$ contains all the two atoms of the setting $s_2$, but $s_3$ includes only one of the two atoms of $s_2$ and $s_{4a}$ only two of the three atoms of $s_3$.
We denote the latter properties with the notation $s_2\subset s_{4a}\subset s_5$ and $s_3\subset s_5$. In addition, we collected also further measurements of settings with 4 atoms. In detail, we run a second setting $s_{4b}$ with the atoms in the same position of $s_{4a}$ and other four settings with different positions for the atoms: $s_{4c}$, $s_{4d}$, $s_{4e}$ and $s_{4f}$.
We choose those specific settings because we want to evaluate the two different scaling: (i) in the number of atoms, (ii) in the number of measurements of different settings with the same number of atoms. Specifically, we consider for (i) $s_2$, $s_3$, $s_{4a}$ and $s_5$ and for (ii) $s_{4a}$,  $s_{4b}$, $s_{4c}$, $s_{4d}$, $s_{4e}$ and $s_{4f}$.

The trained \acp{ann} are composed by a single hidden layer of 100 neurons with ReLU activation function and the output layer with a single neuron with sigmoid activation function. The targets are normalized between 0 and 1 before the training and the inverse transformation is applied to calculate the prediction error. The models are developed in \emph{PyTorch}~\cite{paszke2017automatic,paszke2019pytorch} and trained with mini batch gradient descent to minimize the L1 loss using the Adam optimizer~\cite{kingma2014adam} with learning rate 0.001 and batch size 512. All models are trained with early stopping for a maximum of 150 epochs. 

To perform the scaling (i) we trained four different models using as inputs the $2^2$, $2^3$, $2^4$ and $2^5$ measurements of respectively the settings $s_2$, $s_3$, $s_{4a}$ and $s_5$. To perform the scaling (ii) we consider the measurement coming from the following systems: 
\begin{itemize}
    \item  $s_{4a}$ ($2^4=16$ measurements)
    \item $s_{4a} \oplus s_{4b}$ ($2^4+16=2^5$ measurements)
    \item $s_{4a} \oplus s_{4b} \oplus s_{4c}$ ($2^4+32=48$ measurements)
    \item $s_{4a} \oplus s_{4b} \oplus s_{4c} \oplus s_{4d}$ ($2^4+48=64$ measurements)
    \item $s_{4a} \oplus s_{4b} \oplus s_{4c} \oplus s_{4d} \oplus s_{4e}$ ($2^4+64=80$ measurements)
    \item $s_{4a} \oplus s_{4b} \oplus s_{4c} \oplus s_{4d} \oplus s_{4e} \oplus s_{4f}$ ($2^4+80=96$ measurements).
\end{itemize}

In all the cases, the datasets are split in 20 equal parts to perform a 20-fold cross validation and we report the resulting average mean absolute error and its standard deviation for the 20 models. 

As a remark, the Pulser simulator allows to specify the number of samples per run to speedup the computation. In that case, for each run the final quantum state is preliminary calculated, then the specified number of measurements is obtained from such state. Even if this expedient is useful to spare resources, we found in preliminary experiments that it is counter-productive in the context of noise estimation. In fact, for all the samples of one run, the Hamiltonian defining the evolution is always the same and also the noise that influences it. For this reason, in our work we keep the number of samples per run equal to 1 forcing the resampling of the noise at each single measurement.

Moreover, in this context is better to consider more measurements respect to increase the number of atoms of the setting. In detail, considering both subfigures, the number of data points for the measurement of the setting with 5 atoms in \cref{fig:scaling}(a), i.e. $2^5=32$, is equal to the ones for two concatenated measurements of settings with 4 atoms in \cref{fig:scaling}(b), i.e. $2^4+2^4=32$, but the error in the latter case is lower than the former. 

\subsection{Multiple parameters characterization}\label{sec:multipleParametersCharacterizationAPPENDIX}
Before training the models, the noise parameters are normalised between 0 and 1 to avoid uneven prediction error during the loss calculation. The models are implemented in Pytorch~\cite{paszke2017automatic,paszke2019pytorch} and are trained with minibatch gradient descent to minimise the L1 loss using Adam~\cite{kingma2014adam}.
Regarding the architecture of the models, the \acp{ann} is a \ac{mlp} with the ReLU activation function for all the hidden layers and the sigmoid activation function for the last layer.
The best combination of number of neuron layers, number of neurons in each layer, batch size and learning rate is chosen with an hyper-parameter optimization procedure. The latter is implemented using the python library Ray Tune~\cite{liaw2018tune} with the ASHA scheduler~\cite{li2020system} and the HyperOpt search algorithm~\cite{bergstra2013hyperopt}.
The ASHA scheduler allows multiple models to be trained in parallel, iteratively interrupting the training of the least promising one and thus reducing the duration of the hyper-parameter optimization. In our case at each epoch it halved the models by discarding those with the highest calculated loss on the validation set.
HyperOpt search algorithm, on the other hand, chooses the most probable best combinations of hyper-parameters based on the previously trained and/or stopped models.
By this procedure, the model with the most promising set of hyper-parameters is chosen from 1000 models trained with the Adam optimizer.
The hyper-parameters are sampled in the following ranges: number of hidden layers from 1 to 100, number of neurons in each layer from 5 to 200, batch size in $\{2, 4, 8, 16, 32\}$ and learning rate from $log-uniform(10^{-4}, 10^{-1})$.
At the end, the best hyper-parameters combination is: 1 hidden layer of 117 neurons, batch size 16, initial learning rate $\approx 0.069$, 
dropout probability $\approx 0.044$ 
and L2 regularization $\approx 0.0002$. 

After finding the best set of hyper-parameters, 20 models are trained using the cross validation procedure to exploit the entire dataset and to obtain the standard deviations of the predictions.
In detail, for the cross validation the dataset is divided into 20 equal parts, 18 are used for training, one for validation and one for testing.
The advantage of using the cross validation procedure is that a different block is used for the test of each model and also, in this way all the samples of the dataset are exploited for the training.
Each one of the 20 models is trained with early stopping for a maximum of 150 epochs.

\end{document}